\documentclass[12pt]{article}
\usepackage{graphicx}
\usepackage{amsmath}
\usepackage{amsfonts}
\usepackage{amssymb}

\begin{document}

{\begin{center} \Large {\bf Simulation of hydrogen diffusion and
boron passivation in crystalline silicon}

\normalsize

\hspace{1cm}

O. I. Velichko$^{1*}$, Yu. P. Shaman$^{1,2}$, and A. P.
Kovaliova$^1$

\hspace{1cm}

$^1$Department of Physics, Belarusian State University of
Informatics and Radioelectronics, 6, P.~Brovki Street, Minsk,
220013 Belarus

$^{2}$Scientific-Manufacturing Complex ``Technological Center''
MIET, 5-4806, Street, Zelenograd, Moscow, 124498, Russia

\hspace{1cm}

$^*$Corresponding author, Tel.: +375296998078. {\it E-mail
address:} velichkomail@gmail.com (Oleg Velichko)

\end{center}}

\hspace{2cm}

 \normalsize

{\bf Abstract.} The model of hydrogen migration and of the
reactions of hydrogen atoms with electrically active impurity,
developed earlier, has been applied to simulate hydrogen diffusion
and passivation process during plasma deuteration of silicon
substrates doped with boron. The calculated deuterium
concentration profiles agree well in the length of the passivated
region with the experimental data obtained on treatment in
hydrogen plasma at a temperature of 200 $^{\circ}$C for 5, 10, and
15 minutes. On the other hand, to achieve a good fit to the
abruptness of the calculated profiles between the passivated and
unpassivated regions, it is necessary to suppose that the values
of the parameters that describe the absorption of hydrogen
interstitials by electrically active dopant atoms decrease with
increase in the depth of the passivated region. For example,
nonuniform spatial distributions of nonequilibrium point defects
generated during plasma treatment can lead to a spatial dependence
of hydrogen absorption.

\hspace{1cm}



\section{Introduction}

The increase in the cost of traditional energy resources has been
a worldwide tendency during the past years. Therefore,
investigation and implementation of alternative renewable energy
sources are particularly urgent now. Solar cells provides such
type of promising alternative energy sources. Here, new modules
based on silicon layers constitute a reliable, proven,
sustainable, and environmentally friendly source of energy. It is
worth noting that the best crystalline silicon photovoltaic
modules are 5 $\%$ more efficient than the best modules based on
polysilicon films \cite{Saga-10,Green-12}. It is especially
important for solar cells used in outer space. The treatment of
silicon photovoltaic layers in a hydrogen containing gas plasma
results in a further increase in the solar energy-to-electricity
conversion ratio. All of the above-mentioned factors demonstrate
the importance of elucidating the role of hydrogen in the
evolution of defect-impurity system in the near-surface region of
silicon layers including passivation of dangling bonds,
undesirable defects, and dopant atoms. The main goal of this work
is to analyze theoretical models and carry out calculations of
hydrogen diffusion and hydrogen passivation of electrically active
impurity atoms in silicon crystals doped with boron.

\section{Model}

The model of hydrogen diffusion with account for the passivation
of electrically active impurity atoms was developed in
\cite{Velichko-09,Velichko-10}. It is worth noting that in
contrast to other models of hydrogen diffusion (see, for example,
\cite{Rizk-91}), in the model of Velichko et al.
\cite{Velichko-09,Velichko-10}, as well as in the model of Zhang
\cite{Zhang-02}, the values of hydrogen diffusivity are
approximated from high-temperature experimental diffusivity data
obtained in \cite{Wieringen-56}. Thus, the contradiction between
the diffusivity data obtained for low- and high-temperature
treatments (see \cite{Zhang-02} and references therein) is
eliminated. It is supposed in \cite{Velichko-09,Velichko-10} that
there are two fluxes of hydrogen atoms: ``slow'' diffusing
hydrogen species, responsible for diffusion in the near-surface
region with high a hydrogen concentration, and ``fast'' diffusing
species, responsible for the hydrogen diffusion into the bulk of a
semiconductor in the low-concentration region of the hydrogen
concentration profile. It was supposed that the fast diffusion
occurs due to the long-range migration of nonequilibrium
interstitial hydrogen atoms in singly negatively (H$^{-}$),
neutral (H$^{\times}$), and singly positively (H$^{+}$) charge
states. Due to the high mobility of electrons (holes), the mass
action law is valid for conversions between different charge
states of hydrogen interstitials. The long-range migration of
hydrogen interstitials results in the supersaturation of the bulk
of a semiconductor with hydrogen atoms. Therefore, it is supposed
in \cite{Velichko-09,Velichko-10} that these fast diffusing
species are responsible for the passivation of electrically active
boron atoms due to the following reactions:

\begin{equation} \label{Complex_formation}
 \mathrm{A}^{-} + \mathrm{H}^{ +}  \longrightarrow
 (\mathrm{AH})^{\times} \, ,
\end{equation}

\begin{equation} \label{Complex_formation_add}
 \mathrm{A}^{-} + \mathrm{H}^{\times} \longrightarrow
 (\mathrm{AH})^{\times} + e^{-} \, ,
\end{equation}

\noindent where $\mathrm{A}^{-}$ is the acceptor atom in the
substitutional position; $(\mathrm{AH})^{\times}$ is the neutral
immobile ``boron-hydrogen'' complex, and $\mathrm{e}^{-}$ is the
electron. Then, to calculate the total hydrogen concentration
profile and concentration profile of the electrically active boron
the following system of equations can be used:

1) the conservation law for trapped hydrogen atoms:

\begin{equation}\label{Hydrogen_conservation_law}
\begin{array}{l}
\displaystyle{\frac{\partial \, C^{HTR}}{\partial \, t}}  =
\displaystyle{\frac{k^{HIC}(\chi) \, \,
C^{HI\times}(x,t)}{\tau^{HI}_{i}}} \,

\\
\\
+\,k^{HIA}_{i}k^{HIAC}(\chi) \, C^{HI\times}(x,t) +S^{HT}(x,t) -
G^{HT}(x.t) \, ,
\end{array}
\end{equation}

2) the conservation law for substitutionally dissolved dopant
atoms:

\begin{equation}\label{Dopant_conservation_law}
\displaystyle{\frac{\partial \, C(x,t)}{\partial \, t}}  = -
k^{HIA}_{i}k^{HIAC}(\chi) C \, C^{HI\times}(x,t) \, ,
\end{equation}

3) the stationary diffusion equation for nonequilibrium hydrogen
interstitials:

\begin{equation}\label{Hydrogen_diffusion_equation}
\begin{array}{l}
\displaystyle{\frac{\partial }{\partial x}} \left[ d^{HI}_{i}
d^{HIC}(\chi) \displaystyle{\frac{\partial \,
C^{HI\times}(x,t)}{\partial x}} \right] - \,
\displaystyle{\frac{k^{HIC}(\chi)}{\tau^{HI}_{i}}} \,
C^{HI\times}(x,t) \,
\\
\\
- \, k^{HIA}_{i}k^{HIAC}(\chi) C \, C^{HI\times}(x,t)  +
G^{HI}(x,t) = 0 \, ,
\end{array}
\end{equation}

4 ) the diffusion equation for slow migrating hydrogen species:

\begin{equation}\label{Fick}
\displaystyle{\frac{\partial \, C^{HD}}{\partial \, t}}  =
D^{H}_{i} \displaystyle{\frac{\partial^{\,2} \, C^{HD}}{\partial
\, x^{2}}} + G^{HD} \, ,
\end{equation}

\noindent where

\begin{equation}\label{Diffusivity_concentration_dependence}
d^{HIC}(\chi) =  \frac{\beta^{HI-} \, \chi^{-1} + 1 + \beta^{HI+}
\, \chi}{\beta^{HI-}+1 + \beta^{HI+}} \, \, ,
\end{equation}

\begin{equation}\label{Absorption_hydrogen_concentration_dependence}
k^{HIC}(\chi) =  \frac{\beta^{HIS-} \, \chi^{-1} + 1 +
\beta^{HIS+} \, \chi}{\beta^{HIS-}+1 + \beta^{HIS+}} \, \, ,
\end{equation}

\begin{equation}\label{Absorption_impurity_concentration_dependence}
k^{HIAC}(\chi) =  \frac{1 + \beta^{HIA} \, \chi}{1 + \beta^{HIA}}
\, \, ,
\end{equation}

\begin{equation}\label{Chi}
 \chi = {\frac{{C - C_{B} + \sqrt {\left( {C - C_{B}} \right)^{\,2} + 4n_{i}^{2}}
}}{{2n_{i}} }} \quad {\rm ,}
\end{equation}

\begin{equation}\label{Hydrogen_total}
 C^{HT}=C^{HTR}+C^{HD}+C^{HI} \, \, .
\end{equation}

  .

Here $C^{HT}$ is the total concentration of hydrogen atoms;
$C^{HTR}$ is the total concentration of the hydrogen atoms trapped
by immobile sinks and boron atoms; $C$ and $C_{B}$ stand for the
concentration of electrically active dopant atoms which undergo
passivation and the concentration of charged species with the
opposite type of conductivity, respectively; $C^{HI}$ and
$C^{HI\times}$ represent the total concentration of interstitial
hydrogen atoms and the concentration of hydrogen interstitials in
a neutral charge state, respectively; $C^{HD}$ is the
concentration of the slow diffusing hydrogen species; $\chi$ is
the concentration of charge carriers (holes $p$ or electrons $n$
for passivation of acceptor or donor impurities, respectively)
normalized to the concentration of intrinsic charge carriers
$n_{i}$; $S^{HT}$ and $G^{HT}$ are respectively the rates of
direct trapping and detrapping of hydrogen atoms introduced into
the near surface region of silicon substrate by platelets or other
extended defects; $G^{HI}$ is the generation rate of
nonequilibrium hydrogen interstitials in the surface region due to
the plasma treatment and dissolution of platelets or other defects
which incorporate hydrogen atoms; $G^{HD}$ is the generation rate
of hydrogen atoms participating in slow diffusion; $d^{HI}_{i}$
and $d^{HIC}(\chi)$ are the diffusivity of hydrogen interstitials
in an intrinsic semiconductor and normalized concentration
dependence for diffusivity of this species in a doped
semiconductor, respectively; $D^{HD}_{i}$ is the diffusivity of
the slow migrating hydrogen species;
$\tau^{HI}_{i}=(k^{HI}_{i})^{-1}$ is the average lifetime of
nonequilibrium hydrogen interstitials in an intrinsic
semiconductor; $k^{HI}_{i}$ and $k^{HIC}(\chi)$ are the
coefficient of absorption of hydrogen interstitials in an undoped
semiconductor and the normalized concentration dependence of this
coefficient in a doped semiconductor, respectively; $k^{HIA}_{i}$
and $k^{HIAC}(\chi)$ are the coefficient of absorption of
nonequilibrium hydrogen interstitials, when passivation of
impurity atoms occurs in the near intrinsic silicon, and the
normalized concentration dependence of this coefficient in a doped
semiconductor, respectively.

The empirical parameters $\beta^{HI-}$ and $\beta^{HI+}$ describe
the relative contribution of negatively and positively charged
hydrogen interstitials, respectively, to the total hydrogen
diffusion in comparison with the contribution of neutral
interstitials. The empirical parameters $\beta^{HIS-}$ and
$\beta^{HIS+}$ respectively describe the relative absorption of
negatively and positively charged hydrogen interstitials due to
unsaturated traps in comparison with the absorption of neutral
interstitials. At last, the empirical parameter $\beta^{HIA}$
describes the relative absorption of positively charged hydrogen
interstitials during the passivation of electrically active boron
atoms in comparison with the absorption of neutral interstitials.

It is important to note that the concentration dependences
$d^{HIC}(\chi)$, $k^{HIC}(\chi)$, and $k^{HIAC}(\chi)$ are smooth
and monotone functions of $\chi$ \cite{Velichko-88}. In addition,
these functions have the form traditionally used for the
presentation of effective diffusivity of substitutionally
dissolved dopant atoms in processing simulation codes (see, for
example, \cite{ATHENA-12}). Due to these features, the system of
equations (\ref{Hydrogen_conservation_law}),
(\ref{Dopant_conservation_law}),
(\ref{Hydrogen_diffusion_equation}), and (\ref{Fick}) is very
convenient for numerical solution. It is also worth noting that
the concentration of charge carriers $\chi$ can be calculated
either from the assumption of local charge neutrality (\ref{Chi})
or, more exactly, from the Poisson equation for electrostatic
potential $\varphi$.

\section{Simulation of hydrogen diffusion and boron passivation}

\begin{figure}[!ht]
\centering {
\begin{minipage}[!ht]{11.0 cm}
{\includegraphics[scale=0.9]{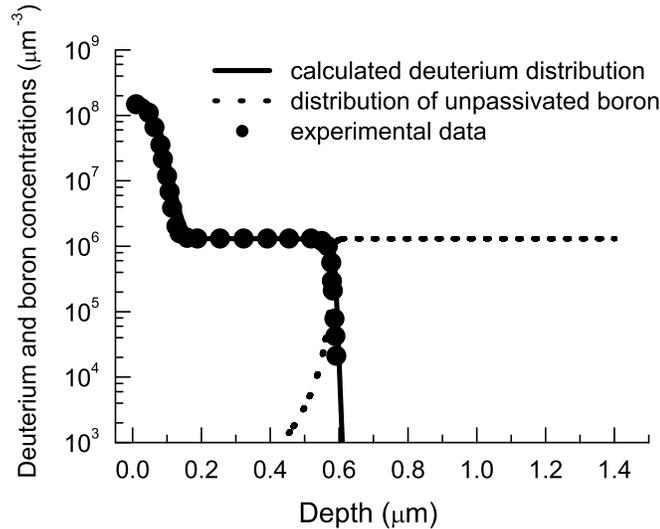}}
\end{minipage}
}

\caption{Calculated profiles of the total deuterium concentration
(solid line) and unpassivated boron concentration (dotted line)
for hydrogenation in the gas discharge plasma at a temperature of
200 $^{\circ}$C for 5 min. The experimental data (black circles)
are taken from Tong et al. \cite{Tong-89}. \label{fig:h51}}
\end{figure}

\begin{figure}[!ht]
\centering {
\begin{minipage}[!ht]{11.0 cm}
{\includegraphics[scale=1.0]{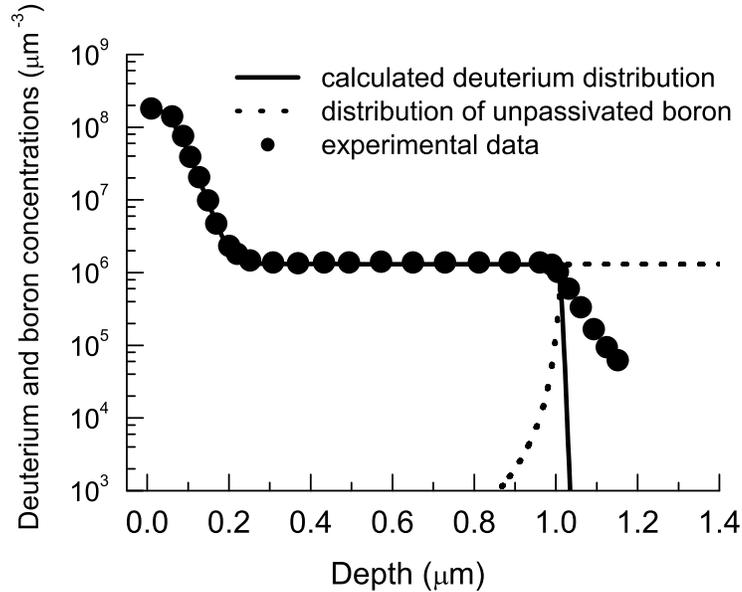}}
\end{minipage}
}

\caption{Calculated profiles of the total deuterium concentration
(solid line) and unpassivated boron concentration (dotted line)
for hydrogenation in the gas discharge plasma at a temperature of
200 $^{\circ}$C for 15 min. The experimental data (black circles)
are taken from Tong et al. \cite{Tong-89}. \label{fig:h152}}
\end{figure}

\begin{figure}[!ht]
\centering {
\begin{minipage}[!ht]{11.0 cm}
{\includegraphics[scale=1.0]{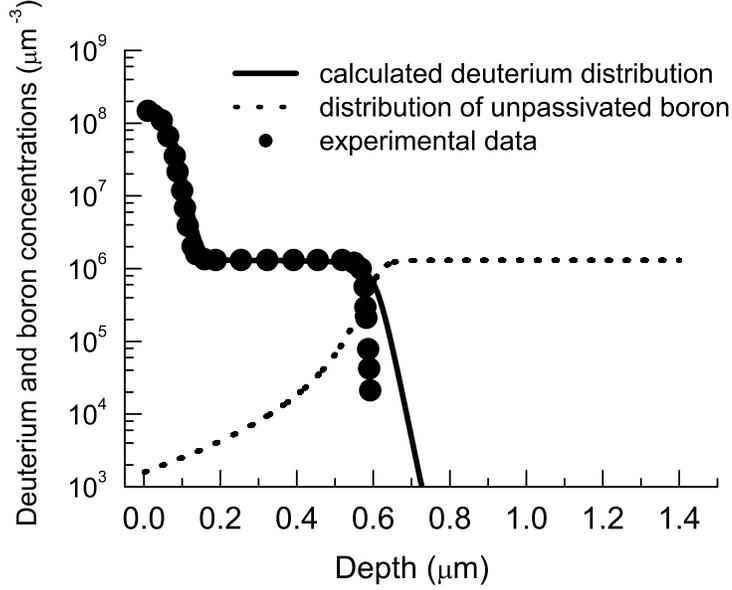}}
\end{minipage}
}

\caption{Calculated profiles of the total deuterium concentration
(solid line) and unpassivated boron concentration (dotted line)
for hydrogenation in the gas discharge plasma at a temperature of
200 $^{\circ}$C for 5 min. The decreased values of $k^{HIA}_{i}$ =
0.2$\times$10$^{-4}$ $\mu$m$^{3}$/s and $\beta^{HIA}$ = 0.12 a. u.
are used in simulation of hydrogen diffusion. The experimental
data (black circles) are taken from Tong et al. \cite{Tong-89}.
\label{fig:h53}}
\end{figure}

\begin{figure}[!ht]
\centering {
\begin{minipage}[!ht]{11.0 cm}
{\includegraphics[scale=1.0]{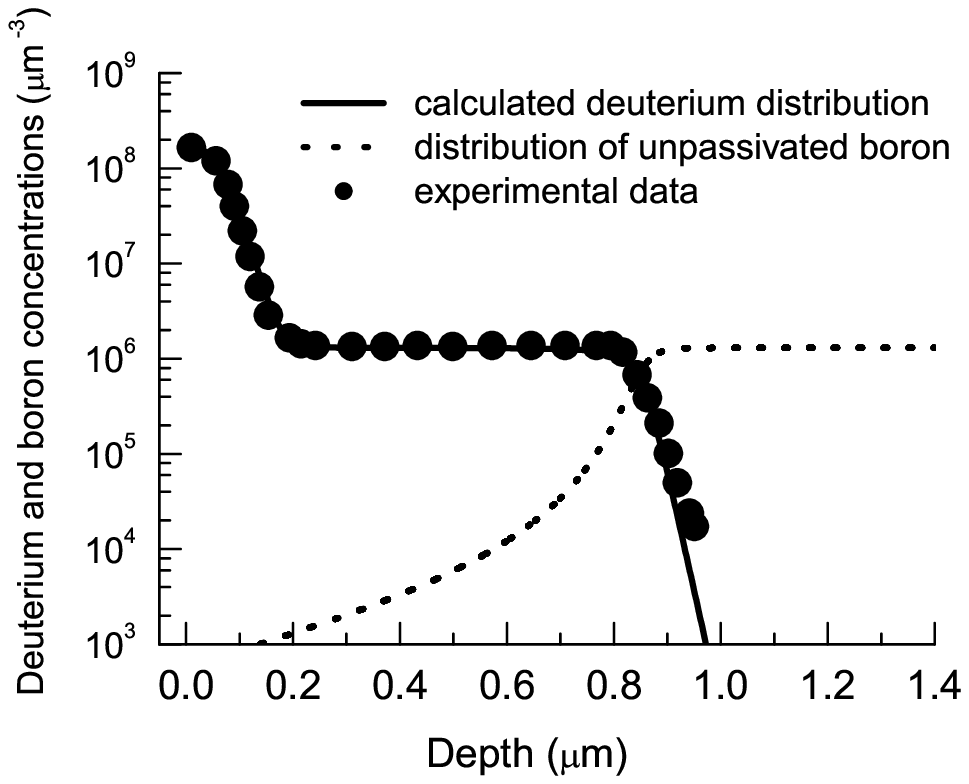}}
\end{minipage}
}

\caption{Calculated profiles of the total deuterium concentration
(solid line) and unpassivated boron concentration (dotted line)
for hydrogenation in the gas discharge plasma at a temperature of
200 $^{\circ}$C for 10 min. The decreased values of $k^{HIA}_{i}$
= 0.2$\times$10$^{-4}$ $\mu$m$^{3}$/s and $\beta^{HIA}$ = 0.12 a.
u. are used in simulation of hydrogen diffusion. The experimental
data (black circles) are taken from Tong et al. \cite{Tong-89}.
\label{fig:h104}}
\end{figure}

\begin{figure}[!ht]
\centering {
\begin{minipage}[!ht]{11.0 cm}
{\includegraphics[scale=1.0]{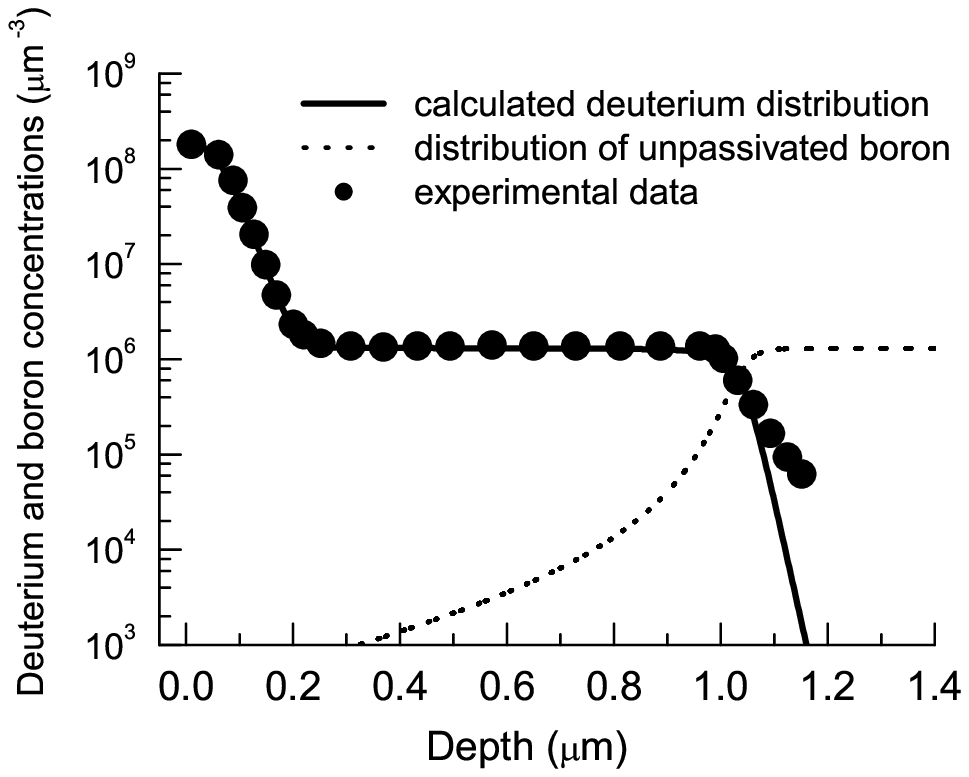}}
\end{minipage}
}

\caption{Calculated profiles of the total deuterium concentration
(solid line) and unpassivated boron concentration (dotted line)
for hydrogenation in the gas discharge plasma at a temperature of
200 $^{\circ}$C for 15 min. The decreased values of $k^{HIA}_{i}$
= 0.2$\times$10$^{-4}$ $\mu$m$^{3}$/s and $\beta^{HIA}$ = 0.12 a.
u. are used in simulation of hydrogen diffusion. The experimental
data (black circles) are taken from Tong et al. \cite{Tong-89}.
\label{fig:h155}}
\end{figure}

To illustrate the efficiency of the model developed in
\cite{Velichko-09,Velichko-10} for describing the hydrogen
diffusion and passivation of boron atoms, the simulation results
for experimental data of Tong et al. \cite{Tong-89} are shown in
Figs.~\ref {fig:h51}, ~\ref {fig:h152}, ~\ref {fig:h53}, ~\ref
{fig:h104}, and ~\ref {fig:h155}. In \cite{Tong-89}, deuteration
was carried out in a theta-pinch plasma or a neutral atom gun.
Deuterium profiling was carried out using SIMS with a cesium-ion
source. The lower limit of the measurable concentration of
deuterium was 10$^{4}$ $\mu$m$^{-3}$. The deuteration temperature
was chosen to be 200 $^{\circ}$C and the surface treatment
durations were 5, 10, and 15 minutes.

To obtain the calculated deuteration concentration presented in
Figs.~\ref {fig:h51}, ~\ref {fig:h152}, ~\ref {fig:h53}, ~\ref
{fig:h104}, and ~\ref {fig:h155}, we used an analytical solution
of Eq. (\ref{Fick}) for the case of the source provides permanent
time-independent generation of a slow diffusing hydrogen species
in the thin surface layer of silicon substrate. Thus, the flux of
the slow component is approximated analytically. The reflecting
boundary condition on the semiconductor surface is imposed on the
low diffusing species with concentration $C^{HD}$. It can be seen
from the experimental profiles that this type of a boundary
condition is in agreement with the shape of hydrogen distribution
in the near surface region. On the other hand, the
finite-difference method \cite{Samarskii-01} is applied to find a
numerical solution for the system of Eqs.
(\ref{Hydrogen_conservation_law}),
(\ref{Dopant_conservation_law}), and
(\ref{Hydrogen_diffusion_equation}). It is supposed that
generation of nonequilibrium interstitials occurs due to the
plasma immersion ion implantation of hydrogen in the thin surface
layer. In addition, the hydrogen interstitials can be generated
due to the rearrangement or dissolution of platelets which
incorporate hydrogen atoms. Unfortunately, the energy of hydrogen
ions was not mentioned in \cite{Tong-89}. However, the thickness
of the layer where generation occurs is negligible in comparison
with the average migration length of hydrogen interstitials and
character dimensions of the passivation region. Therefore, to
obtain a numerical solution, the Dirichlet boundary conditions on
the surface and in the bulk of the semiconductor are imposed on
the rapidly diffusing neutral nonequilibrium deuterium
interstitials with concentration $C^{HI\times}$. Numerical
computations were carried out on a 1D simulation domain
$[0,x_{B}]$, where $x_{B}$ and mesh point number were equal to 1.4
$\mu$m and 7001, respectively. Also, the time grid with 5000 equal
steps was used to obtain a numerical solution.

As can be seen from Fig.~\ref {fig:h51}, the calculated deuterium
concentration profile is in good agreement with the experimental
data of \cite{Tong-89}. The following values of simulation
parameters were used to fit the calculated curve to the
experimental deuterium concentration profile: $Q^{HD}_{i}$ =
1.2$\times$10$^{15}$ cm$^{-2}$; $D^{HD}_{i}$ =
2.0$\times$10$^{-6}$ $\mu$m$^{2}$/s; $d^{HI}_{i}$ = 7.251
$\mu$m$^{2}$/s; $l^{HI}_{i}$ = 4.2 $\mu$m; $C^{HD\times}_{S}$ =
1.02$\times$10$^{2}$ $\mu$m$^{-3}$; $\beta^{HI-}$ = 0;
$\beta^{HI+}$ = 1.0$\times$10$^{-6}$; $\beta^{HIS-}$ = 0;
$\beta^{HIS+}$= 0; $k^{HIA}_{i}$ = 0.21$\times$10$^{-3}$
$\mu$m$^{3}$/s; $\beta^{HIA}_{i}$ = 0.4; $C_{u}$=
1.3$\times$10$^{6}$ $\mu$m$^{-3}$. Here $Q^{HD}_{i}$ is the dose
of hydrogen atoms which are responsible for slow diffusing
species; $C^{HD\times}_{S}$ is the concentration of deuterium
interstitials in the neutral charge state at the surface;
$l^{HI}_{i}$ = $\sqrt{d^{HI}_{i}\tau^{HI}_{i}}$ is the average
migration length of hydrogen interstitials in an intrinsic
silicon; $C_{u}$ is the concentration of uniformly distributed
boron atoms in a substitutional position before hydrogenation.

It follows from the fitting procedure that the increase of
$k^{HIA}_{i}$ or $\beta^{HIA}$ results in a more abrupt deuterium
profile between the passivated and unpassivated regions. The value
of hydrogen diffusivity in intrinsic silicon $d^{HI}_{i}$ = 7.251
$\mu$m$^{2}$/s for a temperature of 200 $^{\circ}$C was taken from
\cite{Wieringen-56}. Because the average migration length of
deuterium interstitials in an intrinsic silicon $l^{HI}_{i}$ is
greater than 1 $\mu$m (it can be seen from Fig.~\ref {fig:h51}
that the thickness of intrinsic silicon to the end of the
passivation process is greater than 0.4 $\mu$m, and the thickness
of hydrogenated layer is approximately equal to 0.6 $\mu$m), we
use in the simulation procedure the value $l^{HI}_{i}$ = 4.2
$\mu$m. Then, the average lifetime of the nonequilibrium hydrogen
interstitials in an intrinsic silicon $\tau^{HI}_{i}$ is equal to
2.43 s, i.e., $\tau^{HI}_{i}$ is significantly smaller than the
duration of hydrogenation (5 min). This value of $\tau^{HI}_{i}$
confirms the correctness of the assumption about the
quasistationary distribution of hydrogen interstitials due to
their high mobility.

In Fig.~\ref {fig:h152}, the calculated deuterium concentration
profile after thermal treatment for 15 minutes is presented. The
same values for diffusion of hydrogen interstitials and the same
passivation parameters were used in these calculations. As can be
seen from Fig.~\ref {fig:h152}, the calculated deuterium
concentration profile is in good agreement with the experimental
data of \cite{Tong-89}, as concerns the length of the passivated
region. On the other hand, the abruptness of the calculated
profile at the boundary between the passivated and unpassivated
regions is very high and disagrees with the experimental data.
Similar calculations of the deuteration distribution for 10
minutes show that in this case the profile abruptness is also
greater in comparison with the experimental one.

It is rather difficult to explain the obtained disagreement at the
boundary between the passivated and unpassivated regions after
plasma treatment for 10 and 15 minutes. Indeed, to reduce the
abruptness of the calculated deuterium profile, one can decrease
the value of $k^{HIA}_{i}$ and/or $\beta^{HIA}$. In Figs.~\ref
{fig:h152}, ~\ref {fig:h104}, and ~\ref {fig:h155}, the calculated
deuterium concentration profiles with $k^{HIA}_{i}$ =
0.2$\times$10$^{-4}$ $\mu$m$^{3}$/s and $\beta^{HIA}$ = 0.12 a. u.
after thermal treatment for 5, 10, and 15 minutes are presented.

It can be seen from Figs.~\ref{fig:h53}, ~\ref {fig:h104}, and
~\ref {fig:h155} that the deuterium concentration profile
calculated for the decreased values of $k^{HIA}_{i}$ and
$\beta^{HIA}$ agree well with the experimental data if plasma
treatment lasts for 10 and 15 minutes whereas for the 5-minutes
duration the slope of the experimental profile is steeper. At
present, we do not know the physical reasons for the decrease in
$k^{HIA}_{i}$ and $\beta^{HIA}$ with increase in the duration of
deuteration and respectively in the depth of the passivated
region. Indeed, the reactions between the hydrogen and boron atoms
are independent of the distance from the surface. Perhaps, there
is an error in the measurements of the deuterium concentration
profile, or the nonequilibrium point defects generated in the near
surface region due to plasma treatment change the conditions for
passivation. For example, generated silicon self-interstitials can
kick-out boron atoms from their substitutional position (Watkins
effect). This problem requires a further investigation.

It is worth noting that simulation of the experimental data of
Tong et al. \cite{Tong-89} was also carried out in
\cite{Zhang-02}. The model of hydrogen diffusion proposed in
\cite{Zhang-02} also takes into account two different fluxes of
hydrogen species. It is supposed that ``fast'' diffusion occurs
due to the migration of atomic hydrogen in different charge states
H$^{-}$, H$^{\times}$, and H$^{+}$. To describe ``slow'' diffusion
in the high concentration region, the mechanism of formation and
migration of hydrogen complexes with mobile traps generated during
plasma treatment is used. It is supposed in \cite{Zhang-02} that
silicon vacancies trap hydrogen atoms and form mobile complexes
V-H, which are responsible for the ``slow'' diffusion. In contrast
to the present paper, only one quasichemical reaction, i.e.,

\begin{equation} \label{Complex_formation_reversible}
 \mathrm{A}^{-} + \mathrm{H}^{ +}  \leftrightarrows
 (\mathrm{AH})^{\times} \, ,
\end{equation}

\noindent is used to describe passivation of boron atoms. It is
also supposed in \cite{Zhang-02} that reaction
(\ref{Complex_formation_reversible}) as distinct from reaction
(\ref{Complex_formation}) is reversible, i. e., the detrapping of
hydrogen atoms can occur. On the other hand, it was found in
\cite{Zundel-89} that the dissociation energy of the (BH) complex
is equal to 1.28 $\pm$0.03 eV, i.e., it is high enough in
comparison with the value of $k_{B}T$ = 0.04077 eV for a
temperature of 200 $^{\circ}$C. Here $k_{B}$ is the Boltzmann
constant. Therefore, in reactions (\ref{Complex_formation}) and
(\ref{Complex_formation_add}) the process of detrapping is
omitted. As can be seen from the paper of Zhang \cite{Zhang-02}, a
full fitting of calculated deuterium concentration profiles to the
experimental ones is not achieved, because the abruptness of the
calculated deuterium distributions at the boundary between the
passivated and unpassivated regions also disagrees with the
experimental data. We hope that new, more precise measurements of
deuterium distributions in silicon substrates doped with boron can
solve this problem.

\section{Conclusions}

The model of hydrogen migration and reactions of hydrogen atoms
with electrically active impurity which had been developed in
\cite{Velichko-09,Velichko-10} was applied for simulating the
hydrogen diffusion and passivation process during plasma
deuteration of silicon substrates uniformly doped with boron. For
comparison, the experimental data of Tong et al. \cite{Tong-89}
for plasma treatment at a temperature of 200 $^{\circ}$C for 5,
10, and 15 minutes was used. The calculated deuterium
concentration profiles agree well with the experimental data for
the entire time of plasma treatment if we suppose that the
coefficient of absorption of hydrogen interstitials in the near
intrinsic semiconductor $k^{HIA}_{i}$ as well the empirical
parameter $\beta^{HIA}$, which describes the relative absorption
of positively charged hydrogen interstitials in comparison with
the absorption of neutral interstitials during the passivation of
electrically active dopant atoms, are decreased with increase in
the duration of deuteration and accordingly in the depth of the
passivated region. It is possible that this decreasing occurs due
to the nonuniform spatial distribution of nonequilibrium point
defects, which are generated during plasma treatment and can
influence the passivation process. If the invariable values of the
simulation parameters are used for different durations of plasma
treatment, the calculated deuterium concentration profile is in
good agreement with the experimental data of \cite{Tong-89}, as
concerns the length of the passivated region. However, the
abruptness of the deuterium concentration profiles at the boundary
between the passivated and unpassivated regions disagrees with the
experimental data either with treatment for 5 minutes or 10 and 15
minutes.

\end{document}